%% file: main.tex
\documentclass[sigconf]{acmart}

\AtBeginDocument{%
  \providecommand\BibTeX{{%
    \normalfont B\kern-0.5em{\scshape i\kern-0.25em b}\kern-0.8em\TeX}}}

\copyrightyear{2024}
\acmYear{2024}
\setcopyright{acmcopyright}
\acmConference[CODS-COMAD 2024]{7th Joint International Conference on Data Science \& Management of Data (11th ACM IKDD CODS and 29th COMAD)}{January 4--7, 2024}{Bangalore, India}
\acmBooktitle{7th Joint International Conference on Data Science \& Management of Data (11th ACM IKDD CODS and 29th COMAD) (CODS-COMAD 2024), January 4--7, 2024, Bangalore, India}\acmDOI{10.1145/3632410.3632487}
\acmISBN{979-8-4007-1634-8/24/01}

\newcommand{\recsys}{\mathtt{RE\mbox{-}RecSys}}
\usepackage{amsmath}

\usepackage{tabularx,booktabs}
\usepackage{graphicx}
\usepackage{multirow}
\usepackage{enumitem}

\begin{document}

\title{$\recsys$: An End-to-End system for recommending properties in Real-Estate domain}

\author{Venkatesh C}
\affiliation{%
  \institution{Housing.com, India}
 \city{}
 \country{}}
\email{venkatesh.c@housing.com}

\author{Harshit Oberoi}
\affiliation{%
  \institution{Housing.com, India}
 \city{}
 \country{}}
\email{harshit.oberoi@housing.com}

\author{Anil Goyal}
\affiliation{%
  \institution{Housing.com, India}
 \city{}
 \country{}}
\email{anil.goyal@housing.com}

\author{Nikhil Sikka}
\affiliation{%
  \institution{Housing.com, India}
 \city{}
 \country{}}
\email{nikhil.sikka@housing.com}

\renewcommand{\shortauthors}{Venkatesh and Oberoi, et al.}

\begin{abstract}
We propose an end-to-end real-estate recommendation system, $\recsys$, which has been productionized in real-world industry setting. 
We categorize any user into $4$ categories based on available historical data: \textit{i)} cold-start users; \textit{ii)} short-term users; \textit{iii)} long-term users; and \textit{iv)} short-long term users. 
For cold-start users, we propose a  novel rule-based engine that is based on the popularity of locality and user preferences. 
For short-term users, we propose to use content-filtering  model which recommends properties based on recent interactions of users. 
For long-term and short-long term users, we propose a novel combination of content and collaborative filtering based approach which can be easily productionized in the real-world scenario. 
Moreover, based on the conversion rate, we have designed a novel weighing scheme for different impressions done by users on the platform for the training of content and collaborative models.
Finally, we show the efficiency of the proposed pipeline, $\recsys$, on a real-world property and click stream dataset collected from leading real-estate platform in India.
We show that the proposed pipeline is deployable in real-world scenario with an average latency of $<$40 ms serving $1000$ rpm.
\end{abstract}




\maketitle

\input{introduction}

\input{algorithm}

\input{experiments}

\input{conclusion}
\bibliographystyle{ACM-Reference-Format}
\bibliography{main}

\end{document}

%% file: introduction.tex
\section{Introduction}
\label{sec:introduction}

Over the past few years, the demand for online real-estate tools has increased drastically due to the ease of access to the Internet, especially in developing countries like India. 
There are many online real-estate platforms for owners, developers, and real-estate brokers to post properties for buying and renting purposes. 
These platforms have more than $3$ million active customers per month  and have more than $1.1$ million active properties.
Considering the number of users and properties, it makes it difficult for users to find relevant properties from a large list of possibilities.
Recommendation Systems (RS) are used to provide personalized recommendations based on historical interactions with the platform. 
The development of an efficient RS is critical from both the company's and the customer's point of view. 
On the one hand, it helps customers to narrow down their choices leading to higher customer satisfaction and retention.
On the other hand, it helps companies to  increase their traffic, conversions, and click-through rates (CTRs).
In this paper, we focus on the real estate recommendation engines where housing selection (for both rent and purchase) is a complex decision-making procedure because people purchase/rent properties infrequently throughout their life.
Typically, users come on an online platform via web/mobile application and expresses their needs using search criteria such as rent/purchase, location, price, locality, number of rooms, etc.
Moreover, they interact with the listed properties on the platform by clicking the property details page (PDP), viewing images, playing property videos, filling customer requirement forms (CRF), dropping leads, etc. 
All these interactions on the platform contribute to user preferences and help RS capture the user preferences for providing relevant recommendations.

 In this work,  we proposed an end-to-end pipeline for \textbf{r}eal-\textbf{e}state \textbf{rec}ommendation \textbf{sys}tem (called as $\recsys$) which we have productionized in real-world industry setting. 
Concretely, we classify any user into $4$ categories: \textit{i)} \textit{cold-start} users with no historical data; \textit{ii)} \textit{short-term} users with recent $10$ minutes of interactions on the platform; \textit{iii)} \textit{long-term} who have more than $10$ minutes of historical data and interacted with at-least $5$ properties on the platform; and \textit{iv)} \textit{short-long term} users who have both historical data as well as recent interactions on the platform.
As we are in an industry setting, it is very crucial to have low latency ($<40$ ms serving $1000$ requests per minute) to have a good customer experience on the platform.
Therefore, we propose to use a combination of rule-based, content, and collaborative filtering for the real-estate recommendation system. 
Specifically, for  \textit{cold-start} users, once users express their needs in search criteria, we recommend those properties with the highest number of leads, user conversions, and property conversions in the chosen locality of a particular city. 
For \textit{short-term} users, we propose to use content-based filtering \cite{lops2011content,bobadilla2013recommender} which  makes use of existing contextual information about the users (e.g. location, price, apartment type) and properties (e.g. apartment details) for a recommendation.
For \textit{long-term} users, we used combination of content and collaborative filtering \cite{chen2018survey} which relies on past interactions and recommends properties to users based on the interactions done by other similar users.
For \textit{short-long term} users, we used a combination of short-term and long-term models. 
We answer the following research questions for real-estate recommendation engines:
\begin{enumerate}[leftmargin=*]
    \item What should be the ideal amount of historical training data required for rent and purchase purposes?
    \item What user impressions/interactions should be considered as implicit feedback for both content and collaborative filtering? 
    % \item What should be an ideal combination of short-term historical data and long-term historical data for \textit{short-long term} users?
    \item How to deploy the solution in a production environment keeping the latency $<40$ ms? \footnote{Demo Video is available at \url{https://www.youtube.com/watch?v=On2JGxACnag}} 
\end{enumerate}

% We extensively validate our proposed pipeline over real-world real-estate datasets and compare the performance with start-of-art models.  
% In the next section, we present the proposed $\recsys$ pipeline followed by experimental results in section \ref{sec:experiments} and conclusion in section \ref{sec:conclusion}.

%% file: algorithm.tex
\section{Proposed Approach }
\label{sec:algo}

Purchasing/renting houses is a complex process as they are expensive and people usually purchase/rent them infrequently. 
Moreover, the behavior and preferences of customers change over time (due to property visits, budget constraints, etc) which adds additional temporal complexity to the housing recommendation problem statement. 
It is important to consider the recent as well as past history of a user while designing a recommendation engine for real estate. 
Therefore, we propose to classify users based on the availability of historical data: cold-start users, short-term users, long-term users, and short-long term users.
In the next subsections, we will explain how we tackled each kind of user and built an end-to-end pipeline. 

% \vspace{-4pt}
\subsubsection{\textbf{Rule-Based engine for Cold-Start Users}}
On real-estate platforms, approximately $25\%$ of users are cold-start users and we have no available historical data for model training. 
As users' short/long history is not available on the platform therefore content and collaborative filtering based methods would not be applicable \cite{rehman2020intelligent}. 
Therefore, we propose to use a rule engine to handle cold start users on the platform and keep the latency of the model in acceptable limits for the production scenarios.

Firstly, for each locality in the city, we build cohorts (in other words, groups), based on locality name, apartment type (e.g. 2 BHK, 3, BHK, etc), profile type (broker and owner), price per square feet bins (bins are created with a gap of $500$K  and $10$k Rupees for purchase and rent respectively) and area bins (created with a gap of 500 sq feet area).
Due to this, we are able to narrow down the search space for a particular search query based on filters. 
Then, we calculate the score for each cohort based on the summation of the following $4$ metrics: \textit{number of flats} (indicative of density); \textit{total leads} (indicative of popularity); \textit{percentage of property conversions} ( another indicative of popularity); and \textit{percentage of user conversions} (indicative of user preference).
Please note that the last $2$ metrics inherently take care of scenarios where the number of properties in a particular cohort may not be high.
Finally, the top-\textit{N} matching cohorts are extracted based on search filters for any user, then we return the randomly chosen top 2 properties from each cohort. 
Intuitively, the random selection of properties from each cohort helps us to overcome the challenge of cold-start problem for new properties.
Each new property will fall under some cohort, randomly choosing properties from each cohort allow us to increase the reach for them which will lead to better customer satisfaction. 
Moreover, it helps to have better learning for content and collaborative filtering models.

\subsubsection{\textbf{Content Filtering for Short-term Users}}
Approximately $20\%$ of users fall under the category of short-term users where we have recent $10$ minutes of interaction data on the platform.
Therefore, it is important to personalize the experience for this category of users which can lead to better conversions and CTRs.
One possible solution is to use collaborative filtering which relies on past interactions and recommend properties based on interactions done by other similar users. 
However, given the number of users and properties, it would not be possible to re-train the collaborative filtering model at the regular interval of $10$ minutes \cite{boutemedjet2012predictive, gu2016learning}. 
Thus, we propose to use content-based filtering approach for short-term users in order to have a better personalization experience in real-time scenario. 
We calculate the similarity scores between the user and properties using a cosine similarity measure where users and properties are represented in the same feature space. 
Then, we rank the properties based on the calculated similarity scores to return the results for a particular search query filter.

For the property vector, we have considered multiple categorical (apartment type and furnishing type)  and numerical (price, built-up area, age, floor number and image count) features  based on the property description.
We convert each feature into a categorical feature and create a binary vector for the property vector.
For area, age, floor number, and image count, we created bins with a distance of $500$ sq feet, $3$ years of age, $2$ floors, and $3$ images, respectively.
For the price, we created bins with a gap of $500$k and $10$k for purchase and rent properties respectively. 
The apartment type (2 BHK, 3 BHK, etc.) and furnishing type (fully furnished, unfurnished, and semi-furnished) are categorical variables kept in their original form.

Users on the platform have different levels of behavior such as clicking, viewing the property details page, dropping a lead using CRF forms, viewing images, etc.
It is important that different behaviors should receive different weights in the user profile vector.
We have divided different activities into $4$ categories based on actions: \textit{i)} ``conversion'' which take into account all the activities where user is planning to submit a lead by filling customer requirement form (CRF); \textit{ii)} ``detail page'' where user is viewing the details for a particular property after opening product detail page; \textit{iii)} ``impressions'' where the user is exploring property by looking into its miscellaneous details; and \textit{iv)} ``other'' activities where user is just scrolling the page and spending some time on rating and other details. 
For any user activity, we calculated the conversion rates from that activity till submitting the CRF form. Based on the conversion rates, we have assigned weights to each action as shown in Table \ref{table:action_weights}.
We have assigned the highest weight to the activity which has the highest conversion rate and maximum business impact. 
Then, we calculate the user profile vector similar to the property vector by converting each feature into the categorical feature as explained previously.
For the user profile vector, we calculate the weighted sum of all the activities done by the user over various properties instead of binary encoding. 
This allows us to capture the user's inclination towards a particular attribute.
Based on the last $10$ minutes of interactions for any user, we calculate the user profile vector and compare it with property vectors in a given locality using cosine similarity.
Finally, we return the top-\textit{N} most similar properties to the user based on similarity scores.

% \begin{table*}[!htbp]
% \caption{Assigned Weightages for different actions performed by the user for both purchase and rent}
% \label{table:action_weights}
% \resizebox{\textwidth}{!}{%
% \begin{tabular}{lcp{2.5cm}p{2.5cm}c}
% \toprule
% \textbf{Category} & \textbf{Actions} & \textbf{Conversion  Rate (Purchase)} & \textbf{Conversion Rate (Rent)} & \textbf{Weightage} \\ \midrule
% \textbf{Conversion} & Submitted CRF & 100 & 100 & 10 \\ \hline
% \textbf{Conversion}  & One time Password Request & 83.5 & 83.1 & 8 \\  \hline
% \textbf{Conversion}  & Open CRF,  Filled CRF & 36.8 & 36.9 & 6 \\  \hline
% \textbf{Detail Page}  & Expand recommendation, video views \& image views & 26.1 & 27.7 & 4 \\  \hline
% \textbf{Impressions}  & locality information,  amenities check, price \& floor plan & 16.8 & 14.7 & 2 \\  \hline
% \textbf{Other} & Rating Check, open PDP \& scrolling & 23.8 & 19.6 & 1 \\ \bottomrule
% \end{tabular}}
% \end{table*}

% \vspace{-10pt}
\begin{table}[]
\caption{Assigned Weightages for different actions performed by the user for both purchase and rent}
\label{table:action_weights}
\vspace{-5pt}
\resizebox{\columnwidth}{!}{%
\begin{tabular}{lcllc}
\hline
\textbf{Category}    & \textbf{Actions}                                                                                          & \textbf{\begin{tabular}[c]{@{}l@{}}Conversion\\ (Purchase)\end{tabular}} & \textbf{\begin{tabular}[c]{@{}l@{}}Conversion\\ (Rent)\end{tabular}} & \textbf{Weight} \\ \hline
\textbf{Conversion}  & Submitted CRF                                                                                             & 100                                                                      & 100                                                                  & 10              \\ \hline
\textbf{Conversion}  & \begin{tabular}[c]{@{}c@{}}One time Password\end{tabular}                                              & 83.5                                                                     & 83.1                                                                 & 8               \\ \hline
\textbf{Conversion}  & \begin{tabular}[c]{@{}c@{}}Open or Filled CRF \\ \end{tabular}                                           & 36.8                                                                     & 36.9                                                                 & 6               \\ \hline
\textbf{Detail Page} & \begin{tabular}[c]{@{}c@{}} recommendation, \\ image/video views\end{tabular}        & 26.1                                                                     & 27.7                                                                 & 4               \\ \hline
\textbf{Impressions} & \begin{tabular}[c]{@{}c@{}}locality info, amenities \\  check, price/floor plan \end{tabular} & 16.8                                                                     & 14.7                                                                 & 2               \\ \hline
\textbf{Other}       & \begin{tabular}[c]{@{}c@{}}Rating Check, \\ open PDP \& scrolling\end{tabular}                         & 23.8                                                                     & 19.6                                                                 & 1               \\ \hline
\end{tabular}
}
\vspace{-18pt}
\end{table}

\subsubsection{\textbf{Collaborative Filtering for Long-term Users}}
\label{sec:long-term-users}
On real-estate platforms, approximately $35\%$ of users have historical data of more than $10$ minutes and have interactions with at least 5 properties on the platform.
For this kind of users, we propose to use collaborative filtering which recommends properties based on the preferences of other similar users.
Since, in the case of real-estate recommendation problem, user-property matrix is highly sparse and it is important to recommend less popular properties as well as well-known properties.
Therefore, we propose to use the matrix factorization technique which factorizes a user-property matrix into two low-ranked matrices, the user-factor matrix and the item-factor matrix that can predict new items for a particular user \cite{koren2009matrix, jannach2010recommender}. 
For each user-property pair in the matrix, we will compute the score based on the interaction done by a user for a particular property as defined in Table \ref{table:action_weights}.
For e.g. for a particular property if a user has submitted the CRF form then we fill the score with value of $10$ or if user have just scrolled the property page then we fill the score with the value of $1$.
This allows us to capture the user preferences at the interaction level and provide more personalized experience.
For matrix factorization, we use Alternating Least Square algorithm\cite{hastie2015matrix} which is implemented in Apache Spark ML and built for large-scale collaborative filtering problems.
The training time for ALS is approximately $30$ minutes for $3$ million and $1.1$ million users and properties respectively. 
Therefore, it is possible to re-train the collaborative model at regular intervals of 1-2 hours.
However, for collaborative filtering, there would not be any significant performance improvement  with $1$-$2$ hours of additional data.
Therefore, we propose to re-train the collaborative filtering model once a day and use content filtering for users who have interacted with more than five properties but were not included in the previous ALS re-training. 
Please note that the content model for long-term users is updated at regular intervals of 2 hours.
Compared to the content model for short-term users (updated at every $10$ minutes intervals), long-term content model have bigger time complexity due to large number of interactions for long-term users.
In production, we have deployed a content+collaborative model for long-term users.

% We train the model at regular intervals of 2 hours due to its long training time. 

\subsubsection{\textbf{Hybrid approach for Short-long term Users}}
In recommendation engine pipelines, it is common to have users with both recent as well as past history on the platform. 
On real-estate platforms, approximately $20\%$ of users fall under this category. 
One solution is to use long-term collaborative filtering model to recommend properties to users. 
However, long-term models are re-trained with a gap of regular intervals of $24$ hours and don't take into account the recent history of the user. 
Therefore, we propose to use a hybrid approach which is a weighted combination of content and collaborative filtering models.
For any user, we take the average of scores from both models and return the top properties accordingly.

\subsubsection{\textbf{$\recsys$ Pipeline}}
\begin{figure}[htbp]
    \centering
    \includegraphics[scale=0.36]{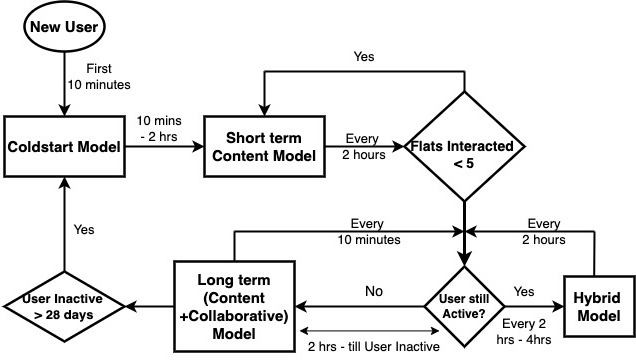}
    \vspace{-10pt}
    \caption{An End-to-End architecture diagram for $\recsys$}
    \label{fig:architecture}
\vspace{-12pt}
\end{figure}

As shown in figure \ref{fig:architecture},  
for any new user, the first $10$ minutes are served using the cold-start rule-based engine. 
If the user is active for  $2$ hours and have interacted with more than $5$ flats, then the user will be served with content (re-trained every 2 hours)+collaborative model (re-trained every $24$ hours) model.
If the user is active for $2-4$ hours, then hybrid model (long-term and short-term models) will return the results to take into account the recent history of user. 
In case of user inactivity for more than 28 days then the user will be served using only cold-start model. 
\vspace{-6pt}

%% file: experiments.tex
\section{Experimental Results and Demo}
\label{sec:experiments}
\subsubsection{\textbf{Dataset}}
We have collected the real-world dataset from our platform which is a leading real-estate online platform in India. 
The platform has an average of $3$ million active users per month and $1.1$ million active properties. 
Moreover, users view $3.57$ pages per visit on average indicating high engagement and interest in the platform.
For the real-estate recommendation system, we need 2 set of information: \textit{i)} property-related data and; \textit{ii)} user interaction events on the platform. 
For the property dataset, we used an internal relational database to collect the information related to properties such as location, price, area of the apartment, property type, etc.
We use Google Analytics to track the user interaction events ($1+$ billion clicks per month) on our platform. 
For our analysis, we have collected the $6$ months of property and event datasets from 1st January to 30th June 2022.

% \subsection{Experimental Setup and Results}
\subsubsection{\textbf{Life Cycle of Buy and Rent Users}} As we are in a real-world scenario where users rent/purchase houses infrequently. 
So, it is important to understand the time taken by the user to make a final decision on the platform. 
Moreover, from a machine learning point of view, this analysis is crucial for setting up the training and testing pipelines. 
From the collected dataset, we analyzed the user persistence in terms of days for both type of users. 
From our experiments, we validated that $85\%$ of rent and buy users persist for approximately $3$ months ($95$ days) and $6$ months ($190$ days) on the platform. 
Therefore, we used the latest $3$ and $6$ months of data for rent and buy users respectively for re-training our models. 

% \begin{figure}[hbt!]
%     \centering
%     \includegraphics[scale=0.2]{User-life-cycle_2.png}
%     %\vspace{-12pt}
%     \caption{Analysis of life cycle for buy and rent users on platform}
%     \label{fig:life_cycle_analysis}
%     %\vspace{-12pt}
% \end{figure}

\subsubsection{\textbf{Train-Test Split}} For collected $6$ months of data, we have randomly sampled click event data consisting of $20,000$ users for each kind of task (cold-start, short-term, long-term and short-long users) separately for rent and purchase use cases. 
In total, we collected $160,000$ user data.
Each user's history is split into train and test data based on random checkpoints. 
By doing this, we are able to test our models on various kinds of users having different amounts of historical data, preferences,  and interactions with the platform. 
Therefore, all the models (content or collaborative) are trained with $20,000$ of training data and are tested with $20,000$ of test data separately for rent and buy. 
 As  followed in literature \cite{bobadilla2013recommender, jannach2010recommender}, we used  MAP@K (Mean Average Precision at K) and NDCG (Normalized Discounted Cumulative Gain) as our evaluation metrics. 

\subsubsection{\textbf{Experimental Results}}
Firstly, we evaluated the performance of the content-filtering model on short-term users. 
Here our objective is to analyze the amount of data required for training the model. 
In table \ref{tab:content_model_results}, we present the results for different iterations of the latest training data i.e. $5,10,20,$ and $30$ minutes. 
It means we train the model with the last $x$ minutes of data and test it on the next $x$ minutes.
From the table, we can deduce that the best results are achieved when we train the model with last $5$ minutes of data.
However, in the production environment, for training the content model, we need to extract the real-time click event data from Google Analytics, pre-process and compute the features. 
This is a time taking process and could not be finished within $5$ minutes given the number of users on the platform. 
Therefore, in production, we used the latest $10$ minutes of data for training the content model for short-term users. 
Moreover, the average inference latency of content model for short-term user is $23.1$ ms ($1000$ requests per minute) which is in acceptable limit in production.

\vspace{-7pt}
\begin{table}[htbp]
\caption{Results for content filtering on short-term users}
\label{tab:content_model_results}
\vspace{-7pt}
\resizebox{\columnwidth}{!}{%
\begin{tabular}{|cl|cc|cc|}
\hline
\multicolumn{2}{|c|}{\textbf{Experiment}}         & \multicolumn{2}{c|}{\textbf{Buy}}                    & \multicolumn{2}{c|}{\textbf{Rent}}                   \\ \hline
\multicolumn{2}{|c|}{\textbf{Train \& Test Sets}} & \multicolumn{1}{c|}{\textbf{MAP@6}} & \textbf{NDCG}  & \multicolumn{1}{c|}{\textbf{MAP@6}} & \textbf{NDCG}  \\ \hline
\multicolumn{2}{|c|}{latest \& next 30 min}       & \multicolumn{1}{c|}{0.853}          & 0.662          & \multicolumn{1}{c|}{0.839}          & 0.625          \\ \hline
\multicolumn{2}{|c|}{latest \& next 20 min}       & \multicolumn{1}{c|}{0.856}          & 0.664          & \multicolumn{1}{c|}{0.845}          & 0.629          \\ \hline
\multicolumn{2}{|c|}{latest \& next 10 min}       & \multicolumn{1}{c|}{\textit{0.866}} & \textit{0.673} & \multicolumn{1}{c|}{\textit{0.856}} & \textit{0.636} \\ \hline
\multicolumn{2}{|c|}{latest \& next 5 min}        & \multicolumn{1}{c|}{\textbf{0.881}} & \textbf{0.69}  & \multicolumn{1}{c|}{\textbf{0.873}} & \textbf{0.646} \\ \hline
\end{tabular}
}
\vspace{-8pt}
\end{table}

Secondly, we have validated the proposed collaborative-filtering model for long-term users trained using the alternating least square algorithm \cite{hastie2015matrix}. 
We have compared the proposed approach (linear weighing of click events) with two other approaches:  
% - \textbf{Exponential Decay:} We use half-life exponential decay for event weights at regular intervals of 3 days. 
% Here, we give more weight to activities that are recent as compared to old impressions.   \\
% - \textbf{TF-IDF Weighing:} We multiply the event weight with inverse property frequency defined as $\log (\frac{\text{total interactions}}{\text{interacations on property}})$.
% This allows us to give more weightage to less popular properties in the corpus. \\

\begin{itemize}[leftmargin=*]
    %\vspace{-2pt}
     \item \textbf{Exponential Decay:} We use half-life exponential decay for event weights at regular intervals of 3 days. 
    Here, we give more weight to activities that are recent as compared to old impressions.  
    \item \textbf{TF-IDF Weighing:} We multiply the event weight with inverse property frequency defined as $\log (\frac{\text{total interactions}}{\text{interacations on property}})$.
    We give more weightage to less popular properties in the corpus.
    %\vspace{-2pt}
\end{itemize}
From results in table \ref{tab:results_long_model}, we can deduce that linear weighing performs best in terms of NDCG for buy and in terms of MAP@6 for rent. For other cases, it is second best compared to baselines. 
In the production environment, the average inference latency for linear weighing collaborative model is $19.29$ ms ($1000$ requests per minute). 
Finally, for \textit{short-long} term users, we use a hybrid approach where we take the average of scores from both models (long-term and short-term) and return the top properties accordingly. 
Moreover, we have evaluated the inference latency for this model and it is within our acceptable limits i.e. $29.3$ ms ($1000$ requests per minute).

% \begin{table}[htbp]
% \caption{Experimental Results on test data for collaborative filtering based model for long-term users}
% %\vspace{-5pt}
% \label{tab:results_long_model}
% \begin{tabular}{|c|l|cc|cc|}
% \hline
% \multirow{2}{*}{\textbf{S.No.}} & \multicolumn{1}{c|} {\multirow{2}{*}{\textbf{Experiment}}}              & \multicolumn{2}{c|}{\textbf{Buy}}                   & \multicolumn{2}{c|}{\textbf{Rent}}                  \\ \cline{3-6} 
%                                 & \multicolumn{1}{c|}{}                                                              & \multicolumn{1}{c|}{\textbf{MAP@6}} & \textbf{NDCG} & \multicolumn{1}{c|}{\textbf{MAP@6}} & \textbf{NDCG} \\ \hline
% 1                               & TF-IDF weighting                       & \multicolumn{1}{c|}{0.713}           & 0.655          & \multicolumn{1}{c|}{0.614}           & \textbf{0.691}          \\ \hline
% 2                               & Exponential decay          & \multicolumn{1}{c|}{\textbf{0.865}}          & 0.662         & \multicolumn{1}{c|}{0.65}           & 0.596         \\ \hline
% 3                               & Linear weighting & \multicolumn{1}{c|}{0.823}          & \textbf{0.685}         & \multicolumn{1}{c|}{\textbf{0.804}}          & 0.646         \\ \hline
% \end{tabular}
% \end{table}
\vspace{-7pt}
\begin{table}[htbp]
\caption{Results  for collaborative filtering for long-term users}
\vspace{-7pt}
\label{tab:results_long_model}
\begin{tabular}{|l|cc|cc|}
\hline
\multicolumn{1}{|c|}{\multirow{2}{*}{\textbf{Experiment}}} & \multicolumn{2}{c|}{\textbf{Buy}}                                     & \multicolumn{2}{c|}{\textbf{Rent}}                                    \\ \cline{2-5} 
\multicolumn{1}{|c|}{}                            & \multicolumn{1}{c|}{\textbf{MAP@6}} & \textbf{NDCG}  & \multicolumn{1}{c|}{\textbf{MAP@6}} & \textbf{NDCG}  \\ \hline
TF-IDF weighting                                  & \multicolumn{1}{c|}{0.713} & 0.655                           & \multicolumn{1}{c|}{0.614} & \textbf{0.691} \\ \hline
Exponential decay                                 & \multicolumn{1}{c|}{0.865} & 0.662                           & \multicolumn{1}{c|}{0.65}  & 0.596                           \\ \hline
Linear weighting                                  & \multicolumn{1}{c|}{0.823} & \textbf{0.685} & \multicolumn{1}{c|}{\textbf{0.804}} & 0.646                           \\ \hline
\end{tabular}
\vspace{-10pt}
\end{table}

\subsubsection{\textbf{REST API and Integration with UI}}
\begin{figure}[htbp]
    \centering
    \vspace{-4pt}
    \includegraphics[scale=0.31]{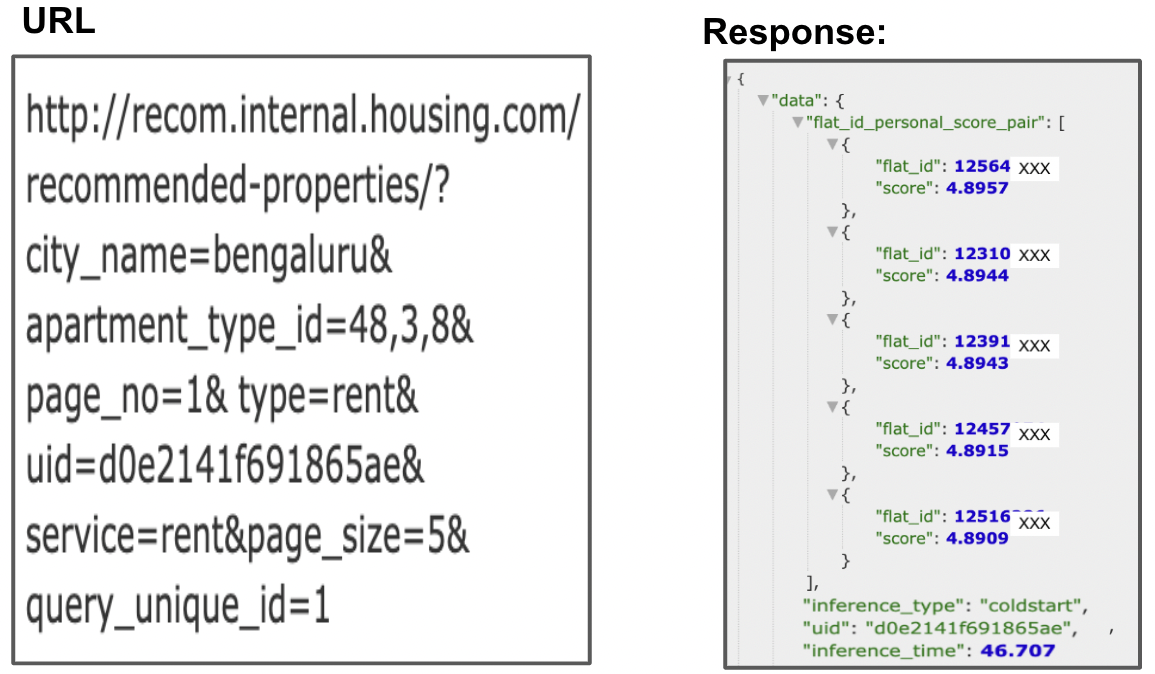}
    \vspace{-10pt}
    \caption{Demo of internal $\recsys$  call Panel}
    \label{fig:demo_call}
\vspace{-8pt}
\end{figure}

\begin{figure}[htbp]
    \centering
    \vspace{-8pt}
    \includegraphics[scale=0.11]{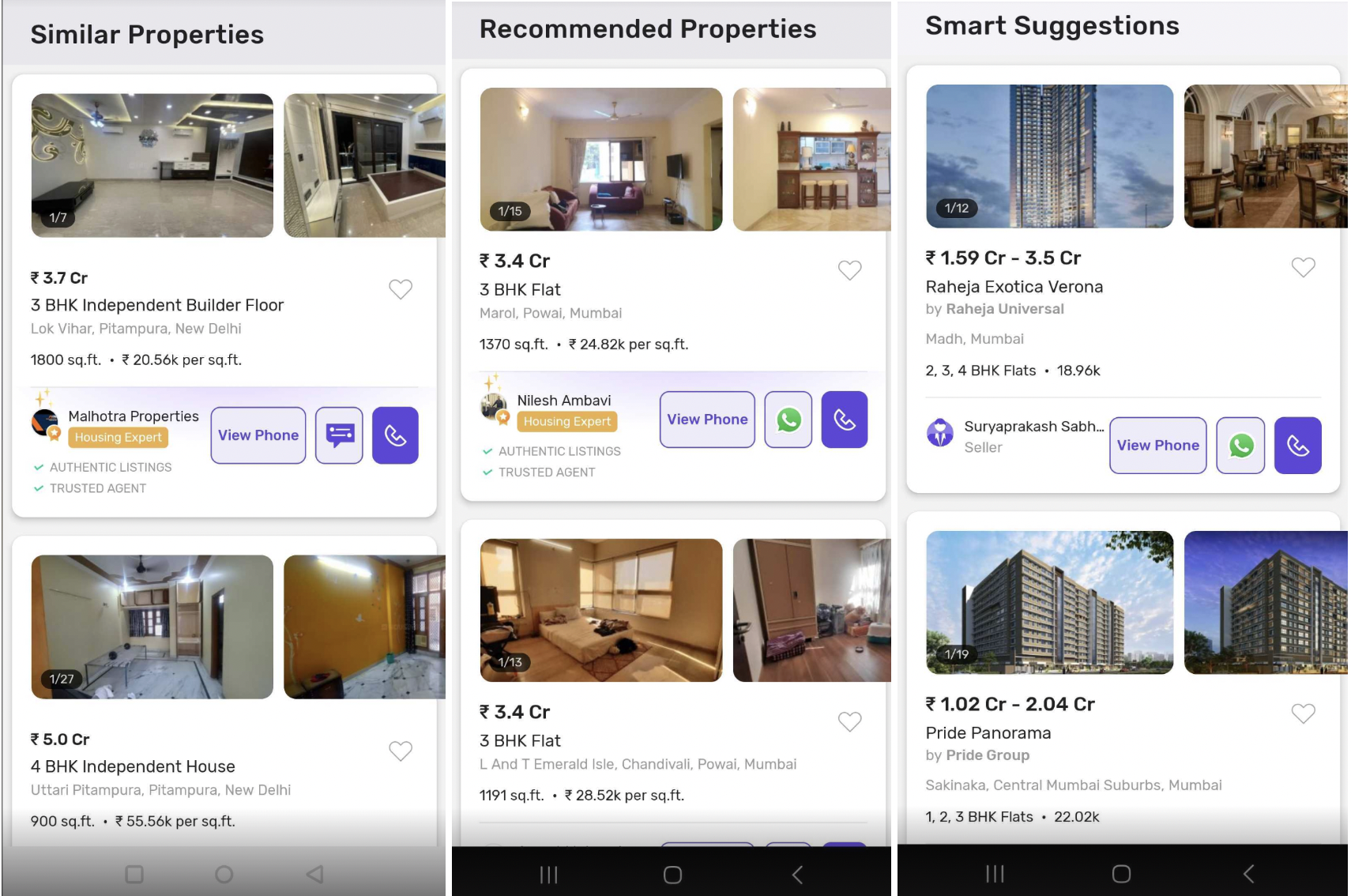}
    % \vspace{-20pt}
    \caption{$\recsys$ integrated with Housing.com App}
    \label{fig:demo_app}
\vspace{-10pt}
\end{figure}

$\recsys$ system is developed in Python and released as REST API.
In figure \ref{fig:demo_call}, we present an example of internal API call along with a JSON response. 
Finally, the results (property ids and corresponding scores) obtained from the API are integrated with various front-end widgets on mobile application, as shown in the figure \ref{fig:demo_app}.
\footnote{Demo Video is available at \url{https://www.youtube.com/watch?v=On2JGxACnag}}

%% file: conclusion.tex
\vspace{-5pt}
\section{Conclusion}
\label{sec:conclusion}
We propose an end-to-end pipeline, $\recsys$, for recommending properties to real-estate users. 
We designed an architecture which can handle different kind of users (cold-start, short-term, long-term, and short-long term users) keeping a balance of infrastructure costs and near real-time personalized recommendations ($<40$ms  serving 1000 rpm).
% For cold-start users, based on user preferences, we designed a rule-based engine that recommends properties from those cohorts (in other words, groups) which are popular and have highest user preference. 
% For short-term (recent $10$ min of interactions ), we propose a content-filtering based solution which is updated at regular intervals of $10$ min in the production environment.
% For long-term users ($>10$ min of interactions), we propose a combination of content and collaborative filtering models which are retrained at regular intervals of $2$ and $24$ hours respectively.
% Finally, for short-long term users where we have both recent $10$ min of interactions and long historical data, we proposed to use a combination of short-term and long-term models. 
% This allowed us to take into account recent history of user along with past interactions for better personalization.
% We have deployed the solution in a real-world environment that has a latency of  less than 40 ms for 1000 rpm. 
We have evaluated the performance of proposed algorithms on a sub-sample of data consisting of interactions from $160,000$ users and compared it with baselines.

%% file: main.bbl

\begin{thebibliography}{9}

%%% the Web address.

\ifx \showCODEN    \undefined \def \showCODEN     #1{\unskip}     \fi
\ifx \showDOI      \undefined \def \showDOI       #1{#1}\fi
\ifx \showISBNx    \undefined \def \showISBNx     #1{\unskip}     \fi
\ifx \showISBNxiii \undefined \def \showISBNxiii  #1{\unskip}     \fi
\ifx \showISSN     \undefined \def \showISSN      #1{\unskip}     \fi
\ifx \showLCCN     \undefined \def \showLCCN      #1{\unskip}     \fi
\ifx \shownote     \undefined \def \shownote      #1{#1}          \fi
\ifx \showarticletitle \undefined \def \showarticletitle #1{#1}   \fi
\ifx \showURL      \undefined \def \showURL       {\relax}        \fi
% invisible to TeX
\providecommand\bibfield[2]{#2}
\providecommand\bibinfo[2]{#2}
\providecommand\natexlab[1]{#1}
\providecommand\showeprint[2][]{arXiv:#2}

\bibitem[Bobadilla et~al\mbox{.}(2013)]%
        {bobadilla2013recommender}
\bibfield{author}{\bibinfo{person}{Jes{\'u}s Bobadilla}, \bibinfo{person}{Fernando Ortega}, \bibinfo{person}{Antonio Hernando}, {and} \bibinfo{person}{Abraham Guti{\'e}rrez}.} \bibinfo{year}{2013}\natexlab{}.
\newblock \showarticletitle{Recommender systems survey}.
\newblock \bibinfo{journal}{\emph{Knowledge-based systems}}  \bibinfo{volume}{46} (\bibinfo{year}{2013}), \bibinfo{pages}{109--132}.
\newblock


\bibitem[Boutemedjet and Ziou(2012)]%
        {boutemedjet2012predictive}
\bibfield{author}{\bibinfo{person}{Sabri Boutemedjet} {and} \bibinfo{person}{Djemel Ziou}.} \bibinfo{year}{2012}\natexlab{}.
\newblock \showarticletitle{Predictive approach for user long-term needs in content-based image suggestion}.
\newblock \bibinfo{journal}{\emph{IEEE transactions on neural networks and learning systems}} \bibinfo{volume}{23}, \bibinfo{number}{8} (\bibinfo{year}{2012}), \bibinfo{pages}{1242--1253}.
\newblock


\bibitem[Chen et~al\mbox{.}(2018)]%
        {chen2018survey}
\bibfield{author}{\bibinfo{person}{Rui Chen}, \bibinfo{person}{Qingyi Hua}, \bibinfo{person}{Yan-Shuo Chang}, \bibinfo{person}{Bo Wang}, \bibinfo{person}{Lei Zhang}, {and} \bibinfo{person}{Xiangjie Kong}.} \bibinfo{year}{2018}\natexlab{}.
\newblock \showarticletitle{A survey of collaborative filtering-based recommender systems: From traditional methods to hybrid methods based on social networks}.
\newblock \bibinfo{journal}{\emph{IEEE Access}}  \bibinfo{volume}{6} (\bibinfo{year}{2018}), \bibinfo{pages}{64301--64320}.
\newblock


\bibitem[Gu et~al\mbox{.}(2016)]%
        {gu2016learning}
\bibfield{author}{\bibinfo{person}{Yupeng Gu}, \bibinfo{person}{Bo Zhao}, \bibinfo{person}{David Hardtke}, {and} \bibinfo{person}{Yizhou Sun}.} \bibinfo{year}{2016}\natexlab{}.
\newblock \showarticletitle{Learning global term weights for content-based recommender systems}. In \bibinfo{booktitle}{\emph{Proceedings of the 25th International Conference on World Wide Web}}. \bibinfo{pages}{391--400}.
\newblock


\bibitem[Hastie et~al\mbox{.}(2015)]%
        {hastie2015matrix}
\bibfield{author}{\bibinfo{person}{Trevor Hastie}, \bibinfo{person}{Rahul Mazumder}, \bibinfo{person}{Jason~D Lee}, {and} \bibinfo{person}{Reza Zadeh}.} \bibinfo{year}{2015}\natexlab{}.
\newblock \showarticletitle{Matrix completion and low-rank SVD via fast alternating least squares}.
\newblock \bibinfo{journal}{\emph{The Journal of Machine Learning Research}} \bibinfo{volume}{16}, \bibinfo{number}{1} (\bibinfo{year}{2015}), \bibinfo{pages}{3367--3402}.
\newblock


\bibitem[Jannach et~al\mbox{.}(2010)]%
        {jannach2010recommender}
\bibfield{author}{\bibinfo{person}{Dietmar Jannach}, \bibinfo{person}{Markus Zanker}, \bibinfo{person}{Alexander Felfernig}, {and} \bibinfo{person}{Gerhard Friedrich}.} \bibinfo{year}{2010}\natexlab{}.
\newblock \bibinfo{booktitle}{\emph{Recommender systems: an introduction}}.
\newblock \bibinfo{publisher}{Cambridge University Press}.
\newblock


\bibitem[Koren et~al\mbox{.}(2009)]%
        {koren2009matrix}
\bibfield{author}{\bibinfo{person}{Yehuda Koren}, \bibinfo{person}{Robert Bell}, {and} \bibinfo{person}{Chris Volinsky}.} \bibinfo{year}{2009}\natexlab{}.
\newblock \showarticletitle{Matrix factorization techniques for recommender systems}.
\newblock \bibinfo{journal}{\emph{Computer}} \bibinfo{volume}{42}, \bibinfo{number}{8} (\bibinfo{year}{2009}), \bibinfo{pages}{30--37}.
\newblock


\bibitem[Lops et~al\mbox{.}(2011)]%
        {lops2011content}
\bibfield{author}{\bibinfo{person}{Pasquale Lops}, \bibinfo{person}{Marco De~Gemmis}, {and} \bibinfo{person}{Giovanni Semeraro}.} \bibinfo{year}{2011}\natexlab{}.
\newblock \showarticletitle{Content-based recommender systems: State of the art and trends}.
\newblock \bibinfo{journal}{\emph{Recommender systems handbook}} (\bibinfo{year}{2011}), \bibinfo{pages}{73--105}.
\newblock


\bibitem[Rehman et~al\mbox{.}(2020)]%
        {rehman2020intelligent}
\bibfield{author}{\bibinfo{person}{Faiza Rehman}, \bibinfo{person}{Hira Masood}, \bibinfo{person}{Adnan Ul-Hasan}, \bibinfo{person}{Raheel Nawaz}, {and} \bibinfo{person}{Faisal Shafait}.} \bibinfo{year}{2020}\natexlab{}.
\newblock \showarticletitle{An intelligent context aware recommender system for real-estate}. In \bibinfo{booktitle}{\emph{Pattern Recognition and Artificial Intelligence: Third Mediterranean Conference, MedPRAI 2019, Istanbul, Turkey, December 22--23, 2019, Proceedings 3}}. Springer, \bibinfo{pages}{177--191}.
\newblock


\end{thebibliography}
